%% file: main.tex
\documentclass[biblatex,english]{lni}
\usepackage{booktabs}

\usepackage[]{blindtext}

\usepackage{multirow}%
\usepackage{algorithm}%
\usepackage{algorithmicx}%
\usepackage{algpseudocode}
\usepackage{wrapfig}

\usepackage{array}
\newcolumntype{L}[1]{>{\raggedright\let\newline\\\arraybackslash}b{#1}}
\newcolumntype{C}[1]{>{\centering\let\newline\\\arraybackslash}b{#1}}

\usepackage{pifont}
\newcommand{\cmark}{\ding{51}}%
\newcommand{\xmark}{\ding{55}}%

\usepackage{todonotes}

\newcommand{\fixSpace}[1]{#1\@\xspace}

\newcommand{\modelRL}{\fixSpace{$M_R$}}
\newcommand{\modelAL}{\fixSpace{$M_A$}}

\newcommand{\subsubsectionsubst}[1]{\textbf{#1.}}

\newcounter{reqcounter}
\newcommand{\req}[2]{\vspace{0.5em}\\\noindent\textbf{Requirement R\refstepcounter{reqcounter}\ref{#1}. \label{#1}}\textit{#2}}

\begin{document}

\title[Multi-Layer Privacy-Preserving Record Linkage with Clerical Review]{Multi-Layer Privacy-Preserving Record Linkage with Clerical Review based on gradual information disclosure}
\author[1]{Florens Rohde}{rohde@informatik.uni-leipzig.de}{0000-0001-7114-1669}
\author[1]{Victor Christen}{christen@informatik.uni-leipzig.de}{0000-0001-7175-7359}
\author[1]{Martin Franke}{franke@informatik.uni-leipzig.de}{0000-0003-4157-8637}
\author[1]{Erhard Rahm}{rahm@informatik.uni-leipzig.de}{0000-0002-2665-1114}

\affil[1]{Leipzig University \& ScaDS.AI Dresden/Leipzig, Leipzig, Germany}
\maketitle

\begin{abstract}
Privacy-Preserving Record linkage (PPRL) is an essential component in data integration tasks of sensitive information. The linkage quality determines the usability of combined datasets and (machine learning) applications based on them.
We present a novel privacy-preserving protocol that integrates clerical review in PPRL using a multi-layer active learning process. Uncertain match candidates are reviewed on several layers by human and non-human oracles to reduce the amount of disclosed information per record and in total. Predictions are propagated back to update previous layers, resulting in an improved linkage performance for non-reviewed candidates as well.
The data owners remain in control of the amount of information they share for each record. Therefore, our approach follows need-to-know and data sovereignty principles.
The experimental evaluation on real-world datasets shows considerable linkage quality improvements with limited labeling effort and privacy risks.
\end{abstract}


\input{1_introduction.tex}
\input{2_related_work.tex}
\input{3_methods.tex}
\input{4_results.tex}
\input{5_discussion.tex}
\input{6_conclusion.tex}

\textbf{Acknowledgments.}
The authors acknowledge the financial support by the Federal Ministry of Education and Research of Germany and by the Sächsische Staatsministerium für Wissenschaft, Kultur und Tourismus for ScaDS.AI.

\newpage
\printbibliography

\end{document}

%% file: 1_introduction.tex
\section{Introduction}

Record linkage, also known as entity resolution, aims at identifying different representations of the same real-world entity, such as a person. It is a crucial step in many data integration tasks in order to combine multiple data sources allowing enhanced data analysis. Typically, unique record identifiers are not available which would enable a join-like operation. Therefore, records are compared pairwise based on their identifying attributes, such as first name, last name and date of birth, and classified as match or non-match.

However, record linkage may potentially harm the privacy of individuals by combining information that can be used against their interests. As a consequence, the conduction of such a linkage is subject to many legal and organizational constraints~\cite{Christen2020springer}.
Privacy-preserving record linkage~(PPRL) methods aim for enabling such linkages without sharing sensitive plaintext information between the data owners or with a third party.
To protect the identifying data, the data owners encode it before sending it to an independent linkage unit which performs the matching on the encoded data only. A variety of such perturbation-based encoding techniques have been proposed, but the most popular and a quasi-standard is based on Bloom filters~\cite{gkoulalas2022tifs}.
An attribute-level application of such techniques results in exploitable frequency patterns in the encoded data. It allows to rather simply reidentify at least some plaintext values, e.g., by aligning the most common last name to the most common Bloom filter value~\cite{Vidanage2022_taxonomy}. The usage of such encodings is thereby limited to data linkages with lower privacy requirements.

The selection and parameter optimization of record linkage approaches typically require training data with information on known matches and non-matches. In practical linkage applications there is generally no such ground truth data available though~\cite{Christen2020springer}.
Linkage on plaintext data can reliably achieve high linkage quality by manually reviewing (uncertain) classifications. Such partial ground truth from clerical review also allows to evaluate and adapt the chosen linkage algorithm. When linking sensitive data, however, a clerical review on plaintext data is usually not feasible.
As a consequence, data custodians may have concerns against the use of PPRL due to its uncertain and potentially lower linkage quality.

There is limited work investigating a privacy-preserving clerical review (PPCR) system for record linkage where attribute values are gradually disclosed and displayed using (visual) masks~\cite{Kum2014,Ragan2018,Kum2019}. However, these masks are applied for display only and the reviewing institution still receives full plaintext data. Moreover, the approach does not aim at reducing the labeling effort or improving an automatic classification model based on labeled samples.

We therefore propose a protocol for PPRL that uses active learning to achieve high-qualitative and reliable linkage results with a low labeling effort. Our linkage protocol employs perturbation-based encodings and uses multiple layers to gradually disclose limited information only if needed. First, the linkage is conducted on record-level encodings where all attributes are combined in a single encoding for each record. These encodings are more secure but do not permit sophisticated classification approaches. Therefore, an active learning process is initiated where uncertain match candidates are iteratively resolved by (re-)classification using attribute-level encodings or ultimately by a masked clerical review. In contrast to a linkage solely based on attribute-level encodings the resistance against reidentification attacks is greatly improved by using pair-specific keys for those encodings and thereby avoiding exploitable frequent bit patterns. 

In particular, we make the following contributions:
\begin{itemize}
    \item We present a novel multi-layer active learning protocol that combines automatic privacy-preserving record linkage and manual masked clerical review while minimizing the amount of shared sensitive data.
    \item We analyze the implications of our protocol with regard to reidentification attacks.
    \item We conduct experiments on real-world data to evaluate the performance of our protocol in terms of labeling effort, linkage quality improvement and privacy risk.
\end{itemize}

%% file: 2_related_work.tex
\section{Background and related work}\label{sec:background-related-work}

In the last decades, a variety of methods for privacy-preserving record linkage has been proposed~\cite{Christen2020springer}.
Some protocols based on secure multiparty computation provide formal security guarantees. However, they typically have very high communication and computing requirements which make them unsuitable for the linkage of large datasets.
Other PPRL methods are based on perturbation techniques where the data owners encode the plaintext, often using some form of cryptographic hashing, before sharing it with a semi-trusted third party for linkage.
The parameters for the encoding, in particular the cryptographic hashing key, are kept secret to the data owners.
Thus, the so-called linkage unit cannot revert the encoding and get access to the plaintext values. This approach is very efficient as it requires low communication costs. The linkage unit can employ blocking techniques to reduce the number of match candidates. In standard blocking, only records that share a certain blocking key, e.g., the same phonetic Soundex code \cite{soundex} of first and last name, are compared.
Such PPRL methods have been used in multiple real-world linkage projects for health research, such as \cite{LUMOS2021, CODEX2022}.
Therefore, we focus on the most popular of those encoding techniques based on Bloom filters. In the following, we describe the technical background from related work and derive requirements for our multi-layer PPRL protocol.

\subsection{Bloom filter based PPRL}\label{sec:bf-based-pprl}
Bloom filter encodings were proposed for PPRL by Schnell et al.~\cite{Schnell2009BF}. They became the de-facto standard for practical PPRL on large datasets due to their straightforward implementation as well as their fast and error-tolerant comparison.
A Bloom filter (BF) is a bit vector of fixed size $m$ where initially all bit positions are set to zero.
The input data is split into overlapping substrings of length $q$ (q-grams). Then, a set of $h$ cryptographic hash functions $\mathbf{H} = \{H_0, H_1, \ldots, H_{h-1}\}$ is applied to each q-gram resulting in bit positions set to '1'. Given that identical q-grams are mapped to the same bit positions, a high overlap of q-grams leads to similar Bloom filters making them suitable for determining the record similarity using set similarity functions, e.g., the Dice coefficient.
This transformation is not reversible due to collisions where multiple features are hashed to the same position.

However, Bloom filter encodings were shown to be susceptible to certain types of attacks~\cite{Vidanage2022_taxonomy}.
Published attacks initially focused on exploiting frequency information of plaintext and encoded attributes as well as pattern mining. 
Recent work uses graph-based attacks to align encoded and plaintext entities by exploiting their similarities to other entities~\cite{VidanageGraph2020}. However, the underlying attack scenario requires an equal or at least very similar plaintext dataset, limiting its practical relevance.
Different encoding techniques to hamper frequency-based attacks have been proposed in the last years. Most importantly, attribute-level encodings that transform each attribute of a record separately should be avoided to prevent an alignment of frequent encoded attribute values to frequent plaintext values~\cite{Christen2018Cryptoanalysis}. 
Instead, multiple or all attributes are combined in a single encoded representation.
Additional hardening techniques can be applied to further distort bit patterns in such record-level Bloom filters~\cite{Franke2021Hard}. A simple but effective approach in terms of utility-privacy trade-off is XOR-folding, where the Bloom filter is split in half and both parts are combined using a bit-wise XOR operation~\cite{Schnell2016Xor}.
\req{req:noabfinfirstlayer}{Attribute-level encodings must not be used with the same parameters for all records to mitigate the risk of successful frequency attacks.}

Unfortunately, such record-level encodings impose limitations that can affect the linkage quality. The encoding parameters are chosen based on assumptions about dataset properties that might be inaccurate. In particular, attribute weights are typically determined by the attributes' value frequencies and error rates. The latter are not known to the data owners prior to the linkage and must be estimated.
Commonly, encoding techniques use fixed weights to ensure that all records are encoded in the same way and are thus comparable. Recent work showed that value-specific weights, e.g., based on the respective value frequency, can be applied in the PPRL context as well, to increase the linkage quality and robustness~\cite{Rohde2023}.
Nevertheless, weighting schemes are still limited for record-level encodings,\input{tables/rbf-issues}in particular since weight adaption cannot be restricted to agreeing attributes as the similarity is not known at the time of the weight application during encoding.
Furthermore, missing values in one record of a pair result in lowered similarity scores even for secondary attributes that could be treated as optional in an attribute-level comparison and classification process.

Typically, the linkage unit computes a single similarity score for each pair of encoded records and classifies it based on a threshold.
The selection of an appropriate threshold is therefore essential for a high linkage quality.
Moreover, record-level encodings conceal whether a certain difference originates from one or multiple attributes. Tab.~\ref{tab:rbf-issues} shows an example where the linkage unit cannot differ between two non-matching records (\texttt{A1} and \texttt{B2}) that have similar but different first names and dates of birth, and two matching records (\texttt{A3} and \texttt{B3}) where these attributes are equal but the city is very different, e.g., because the person moved. A threshold-based classifier that should classify the second pair as a match will therefore misclassify the first pair (and vice versa) due to the equal overall similarity score. 
A classifier with access to attribute-level similarities would be able to distinguish these cases.

\subsection{Privacy-preserving Clerical Review (PPCR)}
Tuning linkage parameters, such as the threshold, requires the availability of (partial) ground truth data which is generally not available in practical linkage projects. Ground truth labels can be determined in a clerical review process where potential matches are decided upon manually.
During clerical review, record pairs and potentially additional information are presented to an oracle, typically a human. The display of such information in plaintext obviously does not preserve privacy and is therefore not applicable to sensitive data. However, systems with a masked display were proposed for manual clerical review that conceal the plaintext by default, present categorical value frequencies, and gradually disclose selected information~\cite{Kum2014,Ragan2018,Kum2019}. The studies showed that the masking had only little impact on the error rate of the labeling using the incremental disclosure approach in \cite{Kum2019} and moderate impact depending on the level of disclosure in \cite{Ragan2018}.
Tab.~\ref{tab:rbf-issues} (bottom) shows an example of such a masked display. Attributes that are identical or very dissimilar are replaced by respective (dis)agreement symbols. Attributes with a medium similarity are displayed partially either by showing the differing plaintext characters (here: first name) or placeholders (here: birth date).
Although the data is shown only partially disclosed, the responsible institution for the clerical review has access to the full plaintext records in the backend services. They are used to determine appropriate masks depending on the attribute similarities.
However, to enable such a privacy-preserving clerical review (PPCR), the system requires merely selected plaintext based on the information whether an attribute pair is equal, dissimilar, or somewhat similar. Only in the latter case, the attribute values are needed to determine replaced or swapped characters (groups).
Based on these observations, we derive the following requirements for our protocol:
\req{req:attr-level-similarities}{The protocol must determine attribute-level similarities.}
\req{req:clerical-review-selected-attributes}{The facility responsible for the (masked) clerical review should only have access to those plaintext attributes that are displayed (partially).}

The risk to conduct a successful re-identification attack on an encoded dataset increases when more information is disclosed (see below in Section \ref{sec:analysis-privacy}).
While some data owners might be willing and allowed to provide more information to improve the linkage quality, others might not. From this assumption, we derive another requirement:
\req{req:optOut}{Data owners must have control over the amount of information they share for each record.}

For example, the provision could be limited by a restrictive or missing consent of some persons. In such a linkage scenario with heterogeneous consents, it would be beneficial to use additional information from records with less strict consent to improve the classification performance for all other records.

\subsection{Active Learning}\label{sec:active-learning}
We aim to use the partial ground truth from clerical review to improve the classification performance for non-reviewed record pairs.
Our protocol is similar to active learning approaches that strive to minimize manual labeling, which in the PPRL context is based on sensitive data. In pool-based active learning, samples are selected from a set of unlabeled instances using a query strategy and labeled by an oracle. The labeled samples are then used to train a classification model. This process is repeated until an exit condition is met, e.g., by reaching a budget of allowed queries.

The most important component is the query strategy. A variety of techniques was proposed~\cite{Papadakis2021}. Some, such as heuristic-based methods using the feature vectors of unlabeled instances, are not suitable for record-level encodings, as only a single similarity score feature is available.
Margin-based strategies are applicable and select the most uncertain instances, typically close to the decision boundary, i.e., the threshold.

While many studies on active learning assume the (human) oracle to be flawless~\cite{Primpeli2020}, this is not a reasonable model for our protocol due to the restricted access to the data for enhanced privacy.
A lack of handling label noise could lead to poor models when used for training. Crowd-based approaches could be used where the output label is determined collectively, e.g., by a majority vote~\cite{Calma2020}. However, in our privacy-sensitive setting, additional queries and oracles would increase the privacy risk. We therefore consider only a single human oracle.
\req{req:noisy-labels}{The model update process must not expect the oracle labels to be error-free as its predictions are based on limited information.}

%% file: tables/rbf-issues.tex
\newcommand{\tabwidth}{0.52\textwidth}
\begin{wrapfigure}{r}{\tabwidth}
\vspace{-0.2cm}
\begin{minipage}{\tabwidth}
\begin{table}[H]
\footnotesize
\centering
\caption{Quality issue of record-level encodings: Matching and non-matching pairs might have the same record similarity score, when they have either multiple slightly different attributes (in \texttt{A1} and \texttt{B2}) or a single replaced attribute (\texttt{A3} and \texttt{B3}). Attribute-level similarities and masked clerical review with limited information disclosure enable better classification.}\label{tab:rbf-issues}
\begin{tabular}{p{0.01\columnwidth}p{0.03\columnwidth}p{0.07\columnwidth}p{0.14\columnwidth}p{0.19\columnwidth}p{0.13\columnwidth}}
\toprule
& \textbf{ID} & \textbf{First} & \textbf{Last} & \textbf{Birth Date} & \textbf{City}\\
\midrule
\multirow{4}{*}{\rotatebox[origin=c]{90}{Plain\footnote{Please note that the full plaintext records are only known to their respective data owners 
and are displayed here in subsequent rows for better comparability by the reader.}}} & \texttt{A1} & \texttt{PAUL\textbf{A}} & \texttt{SMITH} & \texttt{197\textbf{6}/0\textbf{9}/07} & \texttt{RALEIGH} \\
&\texttt{B2} & \texttt{PAUL} & \texttt{SMITH} & \texttt{197\textbf{4}/0\textbf{6}/07} & \texttt{RALEIGH} \\
&\texttt{A3} & \texttt{PETER} & \texttt{COHEN} & \texttt{1976/09/07} & \texttt{\textbf{LELAND}} \\
&\texttt{B3} & \texttt{PETER} & \texttt{COHEN} & \texttt{1976/09/07} & \texttt{\textbf{RALEIGH}} \\
\midrule
\multirow{4}{*}{\rotatebox[origin=c]{90}{Rec.-level}} &\texttt{A1} & \multicolumn{4}{c}{\multirow{2}{*}{0.82}} \\
&\texttt{A3} & \multicolumn{4}{c}{\multirow{2}{*}{0.82}} \\
&\texttt{B3} & \\
\midrule
\multirow{4}{*}{\rotatebox[origin=c]{90}{Attr.-level}} &\texttt{A1} & \multirow{2}{*}{0.8} & \multirow{2}{*}{1.0 \footnotesize (freq.)} & \multirow{2}{*}{0.7} & \multirow{2}{*}{1.0} \\
&\texttt{B2} &  &  &  &  \\
&\texttt{A3} & \multirow{2}{*}{1.0} & \multirow{2}{*}{1.0 \footnotesize (rare)} & \multirow{2}{*}{1.0} & \multirow{2}{*}{0.2} \\
&\texttt{B3} &  &  &  &  \\
\midrule
\multirow{4}{*}{\rotatebox[origin=c]{90}{Masked}} &\texttt{A1} & \texttt{****A} & \multirow{2}{*}{\cmark \footnotesize (freq.)} & \texttt{***\$/*@/**} & \multirow{2}{*}{\cmark} \\
&\texttt{B2} & \texttt{****} & & \texttt{***\%/*\$/**} & \\
&\texttt{A3} & \multirow{2}{*}{\cmark} & \multirow{2}{*}{\cmark \footnotesize (rare)} & \multirow{2}{*}{\cmark} & \multirow{2}{*}{\xmark} \\
&\texttt{B3} & & & & \\
\bottomrule
\end{tabular}
\end{table}
\end{minipage}
\vspace{-0.6cm}
\end{wrapfigure}

%% file: 3_methods.tex
\section{Methodology}

In the following, we present our multi-layer active learning PPRL protocol based on the requirements listed above.

\subsection{Overview}

Our protocol follows the need-to-know principle
: The protocol is comprised of multiple layers of classification with increasing levels of disclosure. The data owners share only as much information for each record as needed for a classification decision with high probability. While the majority of pairs are classified with high certainty in the top layer, some pairs require additional information. The underlying assumption is that the accuracy of the classification benefits from such disclosure.

\begin{wrapfigure}{R}{0.5\textwidth}
\vspace{-0.2cm}
\begin{minipage}{0.5\textwidth}
\begin{figure}[H]
  \includegraphics[width=1\textwidth]{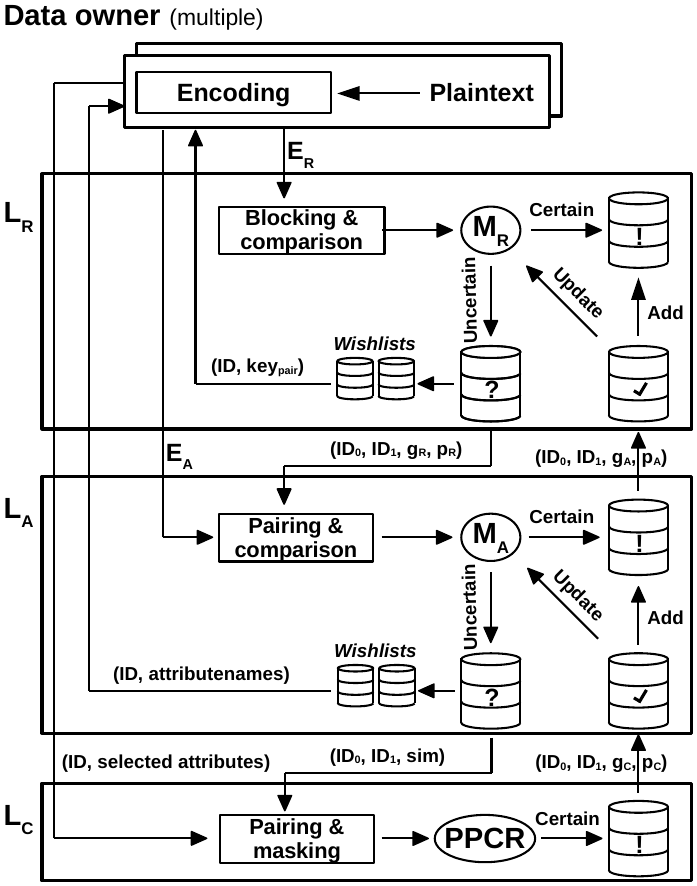}
 \caption{PPRL protocol with three linkage layers: Initially, data owners provide record-level encodings (R). Uncertain pairs are reviewed using keyed attribute-level encodings (A) and finally with masked disclosure of selected attributes only~(C).}
 \label{fig:al-workflow}
\end{figure}
\end{minipage}
\vspace{-0.2cm}
\end{wrapfigure}

Without loss of generality, we focus on a protocol with three layers (see Fig.~\ref{fig:al-workflow}). First, in $L_R$, encodings with a higher focus on privacy protection are used to determine certain matches and certain non-matches, e.g., record-level encodings $E_R$.
Batches of uncertain match candidates are selected for review by the next layer $L_A$. To determine attribute similarities (R\ref{req:attr-level-similarities}) attribute-level encodings $E_A$ are necessary. A pair-specific secret key is used to ensure that each record pair is encoded differently (R\ref{req:noabfinfirstlayer}). These keyed attribute-level encodings serve two purposes: (i) Classify uncertain record pairs with a higher probability for an unambiguous match decision than with record-level encodings and thus ideally make further clerical review unnecessary and (ii) Determine attributes that have an intermediate similarity score and should be visually masked in clerical review (R\ref{req:clerical-review-selected-attributes}). The masked clerical review is the last layer $L_C$ to resolve remaining uncertain match candidates.
For each batch, the revised labels are reported to the upper layer and used to update the model. Non-revised pairs are reclassified with the updated model.
Predictions can change in subsequent iterations and are propagated again upwards in order to hamper learning from erroneous labels (R\ref{req:noisy-labels}).
After each batch, the first layer holds the best possible result that is achieved based on the given information.
A protocol with $i$ layers uses $i-1$ active learning processes.
Our three-layered protocol consists of $\mathit{AL}_R$ and $\mathit{AL}_A$. In $\mathit{AL}_R$, the classifier $M_A$ of layer $L_A$ is used as an oracle to label instances and train the classification model $M_R$ from layer $L_R$. In $\mathit{AL}_A$, a human oracle provides the labels for layer $L_A$ and its model $M_A$.

The protocol is implemented so that data owners actively have to provide additional information via their respective encoding component (R\ref{req:optOut}). They retrieve batches of requests for further information on selected records from the linkage unit. However, the data owners may limit the level of disclosure independently for each record.
If no additional information is provided at all by the data owners to layers $L_A$ and $L_C$, the protocol is equivalent to the common single-provision setting of a PPRL process utilizing a record-level encoding.
If no plaintext attributes are provided, no labeled instances are available for training the attribute-level model. Using a pre-trained model (trained on a sufficiently similar dataset) the protocol could be run in a basic configuration with a single active learning process based on the upper two layers.

\subsection{Components}
\subsubsectionsubst{Encoding}\label{sec:encoding}
Given a record $r$ with a set of attributes $\mathbf{A} = \{a_0, a_1, \ldots, a_{i-1}\}$, we define the encoding function $\mathbf{E} = \textit{encode}(r,\mathbf{P})$, where $\mathbf{P}$ denotes the set of encoding parameters and the output $\mathbf{E}$ is the encoded record consisting of one or more encoded parts.
Record-level encoding functions result in $|\mathbf{E}| = 1$, whereas attribute-level encoding functions produce $|\mathbf{E}| \geq i$ encoded parts, each corresponding to one attribute.
The encoding parameters $\mathbf{P}$ are known only to the data owners (see~Section~\ref{sec:background-related-work}).
Without loss of generality, we focus on Bloom filter encodings, as described in Section~\ref{sec:bf-based-pprl}.
For these encodings the parameters comprise at least the length of the Bloom filter $m$, the q-gram length $q$ and the set of $h$ hash functions $\mathbf{H}$. The hash functions should be independent, e.g., by using Random Hashing~\cite{Niedermeyer2014}. We use a pseudo random number generator (PRNG) to generate $h$ values in the range $[0, m-1]$.
The PRNG is initialized with a seed $s$ that is constructed by a keyed cryptographic hash function $\mathit{HMAC_\mathit{SHA256}}(f, key_{do})\rightarrow~s$, where $f$ is the feature (q-gram) to be hashed and $key_{do}$ is a secret key as part of the parameters $\mathbf{P}$ known only to the data owners.

For the first layer any error-tolerant record-level encoding technique can be used, e.g., record-level Bloom filters as proposed by Durham et al.~\cite{Durham2014} or Cryptographic Longterm Keys (CLK) as proposed by Schnell et al.~\cite{Schnell2011CLK}. Error tolerance means that the encoded entities must be comparable with approximate similarity functions.
For the second layer attribute-level encodings are used, e.g., keyed attribute-level Bloom filter (KABF).
For each attribute $a_i$ a secret key is constructed as $\textit{concat}(key_{do}, key_{pair}, an_i)\rightarrow~key_i$ where $key_{do}$ is the secret key known to the data owners only, $key_{pair}$ is a pair-specific key and $an_i$ is the name of the attribute.
$key_{do}$ is necessary to prevent dictionary attacks by the linkage unit. $key_{pair}$ is a random key generated by the linkage unit individually for each uncertain record pair. It ensures that these specific encoded records are created using the same hash functions and thereby are comparable. The inclusion of the attribute name $an$ is called \textit{attribute salting} and leads to different hash functions for each attribute of a record~\cite{Franke2021Hard}. Thus, the same q-grams from different attributes are hashed to different bit positions which hampers frequency analysis.
The masked clerical review in the third layer requires selected plaintext attributes without any obfuscation. Visual masking depends on the paired records and therefore has to be applied at $L_C$.

\subsubsectionsubst{Blocking, comparison and classification}\label{sec:method-classification}
For each layer a separate linkage strategy is used.
First, pairs of encoded records are generated. We use standard blocking to reduce the number of pairs in the first layer.
For lower layers, no further blocking is required as the match candidates have been identified already.
For each pair the corresponding encoded parts in $\mathbf{E}$ are compared which results in a similarity vector $\mathbf{sim}$.
We use the Dice coefficient to compare Bloom filters and compute normalized similarity scores in the range $[0,1]$~\cite{Dice1945}.
The classification model is a function $\textit{classify}(\mathbf{sim})\rightarrow~(g,p)$, where $g$ is the binary classification target (\{\textit{Non-match}, \textit{Match}\}) and $p$ is the probability of that target ($[0.5,1]$). 
We denote the classification model of layer $L_R$ as \modelRL (Record-level) and the model of $L_A$ as \modelAL (Attribute-level).
Arbitrary classification methods utilizing similarities are applicable.
In the first layer, a simple threshold-based classification model is used. For record-level Bloom filter additional features, e.g., based on the bit vector fillrate, could be computed. Some hardening techniques distort these features though, so for the sake of generalizability we use only the minimal available feature.
In the second layer, we use an evolving Random Forest model as suggested by~\cite{Primpeli2020}. The classifier is updated gradually by adding and replacing trees as explained below.

\subsubsectionsubst{Query strategy}
The aim of the query strategy in Active Learning is the selection of samples that are most important for the training. We consider uncertainty sampling which selects those samples where the probability of the model prediction is low. For a binary threshold-based classification these are typically the record pairs whose similarity is close to the threshold value.
However, a sampling strategy considering only the minimal distance to the threshold potentially results in homogeneous samples for training.
Therefore, we use a bucket-based strategy to select pairs with varying similarities. We divide the samples where $p < p_t$ into $x$ buckets of equal width $(p_t - 0.5)/x$, where $p_t$ is the probability threshold. In multiple iterations, a random sample is selected from each of the buckets ordered by the lower bound of that bin.
For each selected pair we generate a pair-specific secret $key_{pair}$ and push both records to a queue of oracle requests (\textit{wishlist}). According to R\ref{req:optOut} the data owners retrieve a batch of their respective wishlist and may or may not provide the requested record representations to the next layer.
The batch size depends on the overall size of the datasets as well as on the expected response rate of the data owners. A larger batch size is required to gather a sufficient number of pairs if the response rate of the data owners is low (as the data owners' responses are independent of each other).
In $\mathit{AL}_A$, we again apply this query strategy, based on the probabilities $p$ that are calculated by \modelAL.

\subsubsectionsubst{Oracle}
The oracle assigns a (preliminary) ground truth label to a record pair.
In $AL_R$, the attribute-based \modelAL is used as the oracle based on attribute similarities, whereas in $AL_A$ a human assigns the label based on the visually masked display.
Both oracles have limited information and therefore may assign wrong labels with a non-negligible error rate. We denote the error rate of the masked clerical review as $err$.
As the oracle in the intermediate layer $L_A$ evolves with more updates from the lower layer, the oracle may revise its prediction as explained in the next Section.

\subsubsectionsubst{Update and back-propagation}\label{sec:model-update}
Predictions of the oracle are reported from that lower layer and used to update the model. The updated model is then used to reclassify all non-reviewed instances. After that, all instances with changed outcomes $(g,p)$ are reported again to the next upper layer.
The update is based on all labeled instances (also from previous iterations). This is due to the fact that instances may be reported multiple times from the lower layer if the prediction has been revised by the reclassification in the lower layer.

The classifier for record-level encodings \modelRL uses a single threshold $t$. We implement a straightforward threshold optimization algorithm as follows
: The reported labeled pairs with their similarity are classified using various thresholds. The threshold for which the quality measure $q$ is optimized is selected as the new threshold. We consider only thresholds within a maximal distance $dt$ to the initial threshold because we presume an approximately suitable default value. We also restrict the maximal shift per update ($dt_{step}$) in order to prevent selecting thresholds where few labeled samples are available yet.
For attribute-level encodings initially a Random Forest model is bootstrapped based on the predictions from \modelRL. On update, another small temporary Random Forest model is trained based on all labeled instances. The trees are added to the larger Random Forest model. If the number of trees exceeds a limit, the oldest trees are removed. Thereby, the model gradually adapts to the current set of labeled instances and previous potentially erroneous samples are forgotten.
In both layers, the instances are weighted based on their probabilities $p$. The weights of instances labeled by $L_C$ are doubled as they are assumed to be more reliable.

\subsection{Protocol privacy analysis}\label{sec:analysis-privacy}
Perturbation-based PPRL protocols such as ours are commonly based on the Honest-but-curious adversary model~\cite{lindell_2009_secure}. It is assumed that each party follows the protocol but tries to learn as much as possible about the other parties based on the data it receives. Moreover, the linkage unit is assumed to not collude with any data owner. Otherwise a data owner could share the encoding function including its parameters, such a $key_{do}$, with the linkage unit that could transform various possible records $r$, e.g., from a public source, and thereby conduct a dictionary attack to assign individuals to (encoded) records.

\begin{figure}
 \centering
 \includegraphics[width=0.8\textwidth]{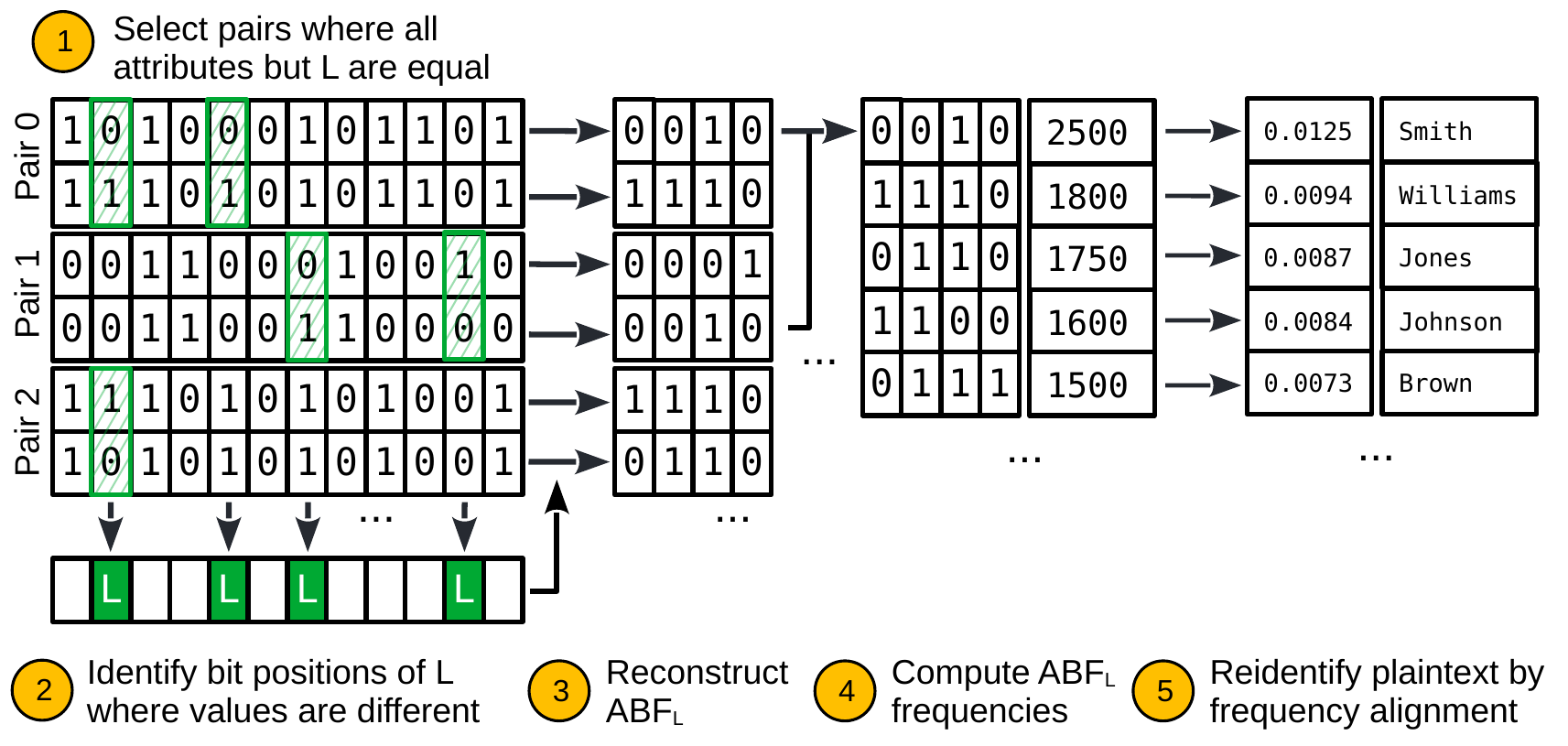}
 \caption{Example of a reidentification attack for an attribute L(astname) based on RBF encodings and known attribute similarities}
 \label{fig:attack-rbf}
\end{figure}

Attribute-level BF have been shown to be vulnerable to frequency attacks~\cite{Christen2018Cryptoanalysis,christen_2018_patternmining}.
These attacks determine frequencies of BF encodings and align these with plaintext value frequencies (see step 4 and 5 in Fig.~\ref{fig:attack-rbf}). A sufficient number of records with the same encoding is required for gathering meaningful frequency information. We therefore use compound secrets where the linkage unit provides distinct secret shares for each record pair as described in Section~\ref{sec:encoding}. Thus, at most two records are encoded the same way and frequencies cannot be determined for the complete dataset.
Nevertheless, attacks based on similarity graphs~\cite{VidanageGraph2020} are still possible in principle. For each attribute a corresponding similarity graph could be constructed. However, as only a subset of the full graph of the first layer is compared on the attribute-level, it is rather unlikely that an attacker could construct a sufficiently similar graph based on plaintext values.

Additional attack scenarios arise in our protocol when multiple linkage layers are conducted by the same linkage unit. In the following, we analyze possible attacks by combining different (encoded) representations of the same records from multiple layers.

\subsubsectionsubst{Attack record-level encodings with KABF}
Durham et al. proposed an encoding method where first attribute-level BF are generated from which bits are sampled according to the respective attribute weights to construct a record-level BF (RBF)~\cite{Durham2014}.
In Fig.~\ref{fig:attack-rbf}, we outline an attack on those RBFs when attribute-level similarities are known. That would be the case if a single organization is responsible for $L_R$ and $L_A$.
This attack is possible because each bit position in the record-level Bloom filter corresponds to exactly one attribute. The bit positions of a certain attribute can be identified using a set of pairs where all but this attribute are equal.

In CLK, tokens from different attributes are hashed directly into a joint Bloom filter and therefore can be mapped to the same bit positions~\cite{Schnell2011CLK}. These collisions have the effect that bit positions cannot be assigned unambiguously to attributes. Thus, the attack on RBF, as described above, is not applicable, in particular if CLK with hardening techniques are used, e.g., with record-specific salting or xor-folding~\cite{Franke2021Hard}.

\subsubsectionsubst{Attack KABF with plaintext}
An attacker with access to a plaintext attribute pair (from $L_C$) and a KABF pair (from $L_A$) can presume which bit positions correspond to which token, similar to the attack on RBFs described above.
However, as distinct hashing secrets are used for each attribute as well as for each record pair, the attacker cannot infer any information about other encoded records.

\subsubsectionsubst{Attack record-level encodings with plaintext}
The most hazardous scenario arises when an attacker has access to plaintext attributes and a record-level encoding of the same record(s) which would be the case when a single (malicious) organization is conducting $L_R$ and $L_C$. Similar to the attacks with KABF above, the attacker may infer correspondences of bit positions/patterns and use this information to attack all other record-level encodings.

To prevent those scenarios, the first layer should be conducted by a different independent organization than the other layers. It receives only the classification outputs of pairs from lower layers.

\textbf{Membership inference.}
Furthermore, the query strategy may leak information to the data owners and allow them to infer the membership of some of their records in other databases.
This could already reveal sensitive information if the other linkage participants and some common characteristics of the records in their databases are known.
For example, the knowledge that there is a duplicate of a known person in a cancer registry leaks private information.
However, the data owners do not learn the classification outcome of their records and thereby cannot tell whether a re-encoding request corresponds to a certain match, a certain non-match or an uncertain match candidate.

%% file: 4_results.tex
\section{Experimental evaluation}
We evaluate our proposed protocol with regard to the linkage quality improvements, the labeling effort and privacy implications.

\subsection{Goals and measures}
\subsubsectionsubst{Quality}
The proposed protocol aims to enhance the overall linkage quality by using labeled instances from clerical review. The improvement is achieved in two ways: (i) The original uncertain labels are replaced by those from lower layers, which are likely more accurate due to the additional available information. (ii) The labeled samples are also used to update the classification models and improve the labels of non-reviewed record pairs by reclassifying them.

To evaluate linkage quality we use the F1 score, which is the harmonic mean of recall and precision. Recall measures the proportion of detected true matches from all true matches. Precision measures the proportion of detected true matches from all detected matches.
The quality assessment is repeated after each batch of reviewed pairs with subsequent post-update reclassification in the first layer.

\subsubsectionsubst{Privacy risk}
We focus on quantifying the privacy risk of lower layers as the top record-level layer is not directly affected by our protocol with regard to feasible attacks.
There is no universal privacy measure for perturbation-based PPRL as the risk depends on the considered attack types and background knowledge as outlined above.
For Bloom filter based encodings several privacy risk scores have been proposed~\cite{Franke2021Hard}.
We report the Gini coefficient (\textit{G}) as well as the Jensen-Shannon divergence (\textit{JSD}) for measuring the dissimilarity of the bit frequency distribution with a uniform distribution. The notion of these measures is that the risk of frequency attacks is reduced if all bit positions have the same likelihood of being set to '1'. The scores of both measures range from $0$ (identical, low privacy risk) to $1$ (maximal different, highest privacy risk).
As the chances of successful attacks rise in general with more accessible information, we also consider the number of available pairs/records in $L_A$ and $L_C$.

For the lowest layer with masked clerical review, we report the share of attributes that have been provided (on request) by the data owners as some attributes are more relevant for the privacy risk than others. However, reidentifications are mostly feasible using attribute combinations as this may allow to unambiguously map partial records with (uncommon) values to their original representation. We therefore also report the k-Anonymized Privacy Risk (KAPR) score which has been proposed for measuring the privacy risk based on the revealed information in masked display~\cite{LiKAPR2019}. The normalized risk score in $[0,1]$ is higher, the more plaintext of the records is disclosed and the lower the number of records that are indistinguishable based on the level of disclosure.
For each record $i$ the number of possible records ($k_i$) is determined based on the available information. The overall KAPR score is computed using an adapted function from \cite{LiKAPR2019}: $KAPR = \frac{1}{ND}\sum_{i=0}^{2n-1} \frac{d_i}{k_i}$ where $N$ is the total number of records (in this layer), $D$ is the total number of attributes and $d_i$ is the number of provided attributes of record $i$ in this layer.
The original formula includes the proportion of disclosed characters, but our KAPR variant measures the risk based on the data available to the reviewing institution instead of based on the information that is displayed.

\subsection{Setup}
\subsubsectionsubst{Datasets}
We use personal records from the North Carolina voter register (NCVR) as provided by Panse et al.~\cite{Panse2021}.
The database contains multiple snapshots of the register and thereby real-world errors of matching records, e.g., due to people moving or changing their names. Ground truth data is available based on unique voter IDs. We use the attributes \textit{first name} (FN), \textit{middle name} (MN), \textit{last name} (LN), \textit{year of birth} (YOB), CITY, ZIP \textit{code} and \textit{place of birth} (POB) for linkage.
We derive multiple datasets with two sources with $50k$ records each that overlap by $10\%$ (S), $20\%$ (M) and $30\%$ (L).
The duplicates are selected by randomly sampling records from snapshot '2021-01-01' and using a duplicate from a different snapshot where at least one (\textit{E1}) or two (\textit{E2}) of the attributes FN, MN, LN, POB, (CITY+ZIP) are non-equal. Our datasets therefore do not contain any perfect matches as those are trivial to match.
Apart from that, the records are not synthetically modified.
These dimensions of dataset variation -- overlap of data sources and disparity of duplicates -- have been chosen as they represent dataset characteristics that typically are not known prior to linkage.
The datasets names are composed of their error rate (E1, E2) and overlap (S, M, L).

\subsubsectionsubst{Encoding}
We use CLK encodings with $m = 1024$ and $h = 12$ for the first layer ($E_R$). Additionally, we test an encoding variant where the XOR-folding hardening technique is applied to these CLK.
As attribute-level encodings $E_A$ we use KABF with $m = 256$ and attribute-specific $h$ (see Table~\ref{tab:abf-k}), following the encoding procedure described in~\cite{Rohde2023}.
\begin{wrapfigure}{R}{0.45\textwidth}
\vspace{-0.4cm}
\begin{minipage}{0.45\textwidth}
\begin{table}[H]
  \caption{Number of hash functions \textbf{h} for the Keyed Attribute-level Bloom filter encodings
  }
  \label{tab:abf-k}
  \begin{tabular}{L{0.02\columnwidth}C{0.03\columnwidth}C{0.05\columnwidth}C{0.04\columnwidth}C{0.06\columnwidth}C{0.08\columnwidth}C{0.04\columnwidth}C{0.06\columnwidth}}
    \toprule
    & FN & MN & LN & YOB & CITY & ZIP & POB\\
    \midrule
    \textbf{h} & $18$ & $21$ & $17$& $26$ & $13$ & $21$ & $43$\\
  \bottomrule
\end{tabular}
\end{table}
\end{minipage}
\end{wrapfigure}

Both, CLK and KABF, have an approximate average fill rate (proportion of 1-bits) of $40\%$.
For $L_A$ and $L_C$, data owners also provide approximate frequency information for the attributes, together with the attribute-level encodings. Possible values are: '1' if the value is in the top $1\%$ most frequent values, '2' in top $5\%$ and '3' if it is rarer. Those frequency labels are used as additional features for \modelAL when the corresponding attributes of a pair are equal, otherwise the value is set to '0'.

\subsubsectionsubst{Linkage}
We conduct the following blocking strategy in $L_R$ to generate candidate record pairs:
For each record we derive multiple blocking keys at the data owners based on the plaintext attribute combinations FN+YOB, LN+YOB and Soundex(FN)+Soundex(LN) and encode each of them using a cryptographic one-way hash function. These hashed blocking keys are transmitted together with the encoded records to enable standard blocking at the linkage unit.
This procedure ensures that the same candidate pairs are generated during the initial batch matching on the first layer even with different encoding techniques.

\subsubsectionsubst{Query strategy and models}
We select the parameters $p_t = 0.8$ and $x = 10$ for our uncertainty-bucket-based query strategy. The prediction probability $p$ of \modelRL is computed for a similarity $sim$ and the threshold $t$ as $0.5 * (1 + min(1, abs(sim - t)/d) \rightarrow~p$ with $d=0.05$ for $sim<t$ and $d=0.1$ for $sim>=t$.
For \modelRL, we apply the threshold shift approach (see Section \ref{sec:model-update}) with accuracy as the quality measure.
For \modelAL, we use the shifting Random Forest classifier based on the RF implementation from the WEKA library~\cite{frank2016weka} using the options \texttt{\small"-P 70 -I 10 -J 10 -N 100 -num-slots 1 -K 0 -M 1.0 -V 0.001 -S 1 -depth 6"} where \texttt{\small -I, -J, -N} determine the initial, added and maximum number of trees. The model is bootstrapped using the initial batch of prelabeled pairs from $L_R$. In each iteration the model is updated with all samples and their respective predictions, either from $L_R$ or $L_C$.

\subsubsectionsubst{Test protocol}
Initially, a batch matching on the first layer is conducted.
Afterwards, our iterative protocol is composed of two phases:
First, $5$ smaller batches of $100$ uncertain record pairs from the first layer are selected for review by the second layer. After classification based on attribute-level encodings with \modelAL, $2$ batches of uncertain pairs on that layer are reviewed by the final oracle (PPCR).
After each reviewed batch, the model is updated based on the revised predictions and all non-reviewed record pairs are reclassified based on the revised model.
The batch size in $\mathit{AL}_A$ is computed as $\frac{b}{10}$.
After this warm-up phase, the budget is reached and the clerical review layer is omitted. Consequently, $M_A$ is not updated anymore. The batch size is increased to $1000$ pairs per iteration for $4$ additional batches to study the performance of the trained $M_A$.

\subsubsectionsubst{Evaluation procedure}
We evaluate different scenarios with regard to the PPCR layer: We vary the simulated error rate $err = \{0.0, 0.1, 0.2\}$ and the clerical review budget $b = \{100, 200, 300\}$. 
For each dataset, we determine the top-level threshold $t_{opt}$ that optimizes the F1 score given the global ground truth. Each scenario is evaluated using the initial thresholds 
$\textsc{range}(t_{opt} - 0.05, t_{opt} + 0.05, 0.01)$
to study the performance of the protocol with regard to threshold optimization.
We repeat each experiment three times and report the micro average F1 score as well as the minimum-maximum range.

\subsubsectionsubst{Baselines}
In the following figures, grey-dotted vertical lines mark the end of the warm-up phase and green dashed horizontal lines represent the F1 score with optimal threshold for the respective dataset without any revised labels as a baseline.
Furthermore, orange-doted horizontal lines refer to the optimal F1 scores achieved using a linkage strategy solely based on attribute-level encodings. Attribute similarities are aggregated to a record similarity using a weighted mean average.
\begin{wrapfigure}{r}{0.5\textwidth}
\vspace{-0.2cm}
\begin{minipage}{0.5\textwidth}
\begin{table}[H]
  \caption{Attribute weights \textbf{w} used for the baseline linkage solely based on (non-keyed) attribute-level Bloom filters.}
  \label{tab:abf-w}
  \begin{tabular}{C{0.015\columnwidth}C{0.06\columnwidth}C{0.06\columnwidth}C{0.06\columnwidth}C{0.06\columnwidth}C{0.08\columnwidth}C{0.06\columnwidth}C{0.06\columnwidth}}
    \toprule
    & FN & MN & LN & YOB & CITY & ZIP & POB\\
    \midrule
    \textbf{w} & $12.04$ & $15.15$ & $5.12$& $6.58$ & $8.23$ & $10.95$ & $6.63$\\
  \bottomrule
\end{tabular}
\end{table}
\end{minipage}
\vspace{-0.4cm}
\end{wrapfigure}
Weights are determined based on a probabilistic approach~\cite{Fellegi1969},
using the default average value frequencies and error rates of the Epilink Matcher based on a German cancer registry~\cite{Rohde2021}. For missing attributes, the corresponding similarity scores are excluded from the aggregation.

\subsection{Results}
\subsubsectionsubst{Dataset E1M}
Fig.~\ref{fig:quality-2232-eX-bX} (left) shows the results for dataset E1M with $b = 100$ and varying $err$.
The worst initial F1 score is $0.791$ for the threshold $t_{opt} + 0.05$. The F1 score for $t_{opt}$ is $0.863$ and the average initial F1 score over all thresholds is $0.838$.
As expected, the linkage quality improves with growing number of reviewed pairs.
After reviewing 4500 pairs on the top layer, average F1 scores reach $0.899$ ($err=0.0$), $0.888$ ($err=0.1$) and $0.881$ ($err=0.2$).
The final results of the best runs are very similar with a F1 score of $0.910\pm0.002$.
The worst runs lead to F1 scores between $0.832$ ($err=0.2$) and $0.863$ ($err=0.0$).
Hence, higher error rates of the PPCR oracle lead to higher outcome variability.

\def\lqplotwith{0.48\linewidth}
\begin{figure}[ht]
\begin{minipage}[b]{\lqplotwith}
\centering
\includegraphics[width=\textwidth]{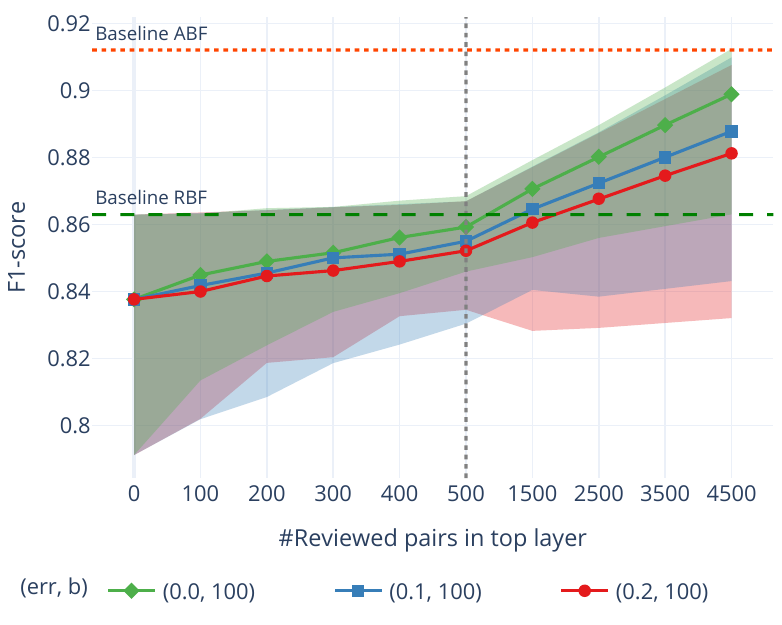}
\end{minipage}
\hspace{0.5cm}
\begin{minipage}[b]{\lqplotwith}
\centering
\includegraphics[width=\textwidth]{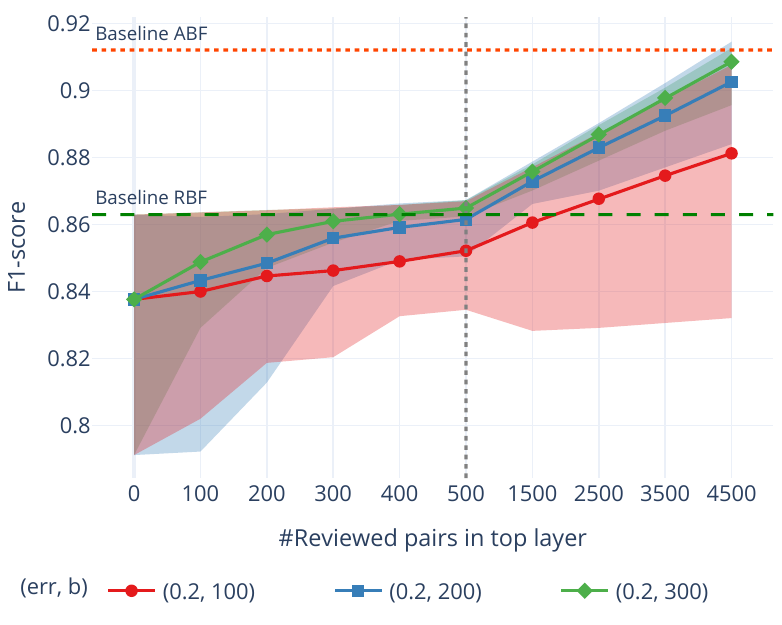}
\end{minipage}
\caption{Linkage quality development for dataset E1M with fixed budget $b=100$ and varying error rates (left) and varying budget $b$ and fixed error rate of $err = 0.2$ (right).}
\label{fig:quality-2232-eX-bX}
\end{figure}

\begin{figure}[ht]
\begin{minipage}[b]{\lqplotwith}
\centering
 \includegraphics[width=\textwidth]{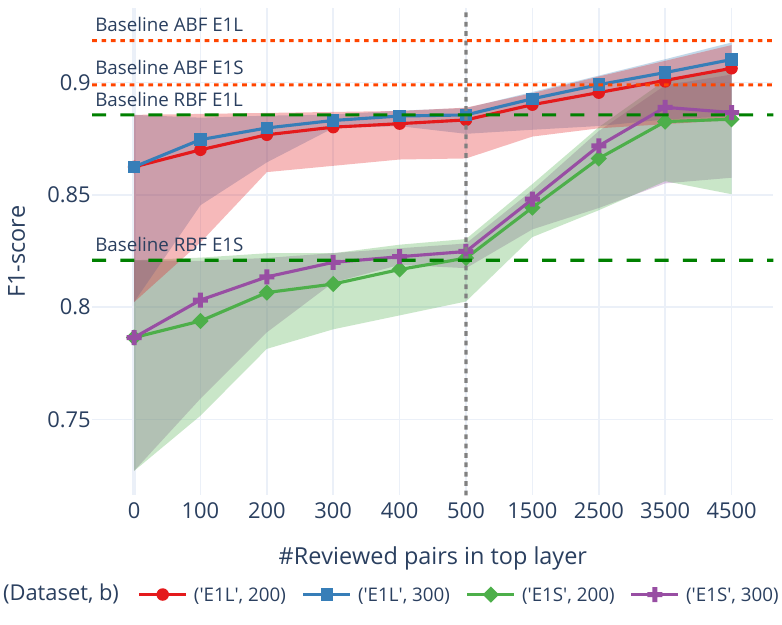}
\end{minipage}
\hspace{0.5cm}
\begin{minipage}[b]{\lqplotwith}
\centering
 \includegraphics[width=\textwidth]{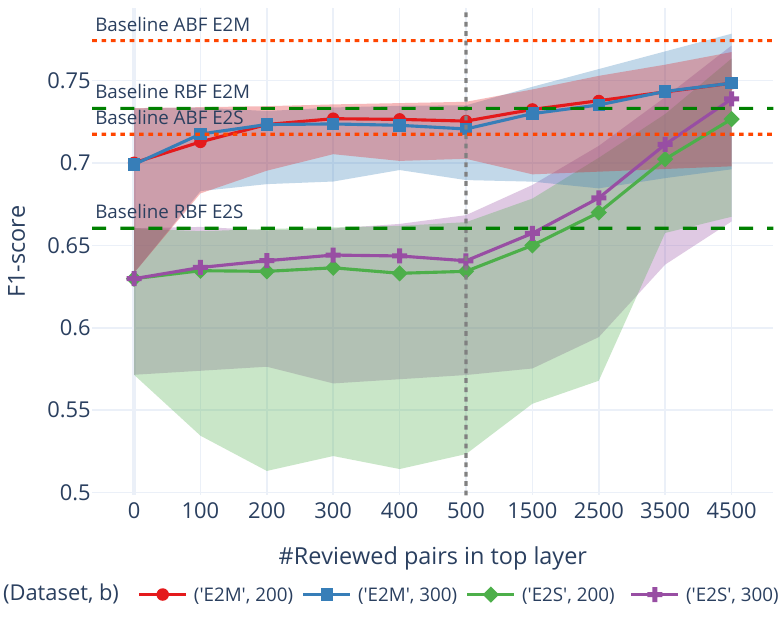}
\end{minipage}
\caption{Linkage quality development for datasets E1S and E1L (left) and E2S and E2M (right) with varying budget $b$ and fixed error rate of $err=0.2$.}
 \label{fig:quality-22xx-bX}
\end{figure}

In the following, we set the error rate $err$ to $0.2$ which is in line with the empirical studies based on visual disclosure in \cite{Ragan2018, Kum2019}.
As depicted in Fig.~\ref{fig:quality-2232-eX-bX} (right), higher budgets $b=200$ and $b=300$ improve the performance considerably. The final average F1 scores are $0.903$ ($+0.065$) and $0.909$ ($+0.071$) which is comparable to the best runs with lower budget.
Higher budgets also improve the reliability of the results, as even the worst runs achieve F1 scores of $0.884$ and $0.896$. The range of outcomes is reduced from $0.072$ to $0.031$ ($b=200$) and $0.017$ ($b=300$).
Furthermore, F1 scores rise already with a low number of reviews in the top layer. In our setup, the optimal threshold can be reached from the worst starting condition ($t_{opt}\pm 0.05$) after three iterations due to the shift limit $dt_{upd}=0.02$. For $b=300$ the average F1 score reaches the baseline in three to four rounds ($90-120$ manual reviews) whereas for $b=100$ it is not achieved at all during the warm-up phase despite a comparable number of manual reviews. Therefore, the quality improvements with larger budgets in that phase are mostly the result of the threshold shift algorithm.

\subsubsectionsubst{Datasets E1S and E1L}
Experiments using dataset variants with lower and higher overlap show enhancements of the average F1 scores 
from $0.786$
by $0.101$ (E1S, $b=300$) and
from $0.862$ by
$0.048$ (E1L, $b=300$), see Fig.~\ref{fig:quality-22xx-bX} (left).
The minimal F1 score is increased from $0.727$ by $0.131$ (E1S, $b=300$)
and from $0.802$ by $0.083$ (E1L, $b=300$).
The range of outcomes is reduced from $0.094$ to $0.046$ (E1S, $b=300$) and from $0.083$ to $0.033$ (E1L, $b=300$).
The final results for $b=200$ are comparable, but the improvements are achieved later and with higher variance in the process.

\begin{wrapfigure}{r}{0.50\textwidth}
\vspace{-0.5cm}
\begin{minipage}{0.50\textwidth}
\begin{figure}[H]
 \centering
 \includegraphics[width=\textwidth]{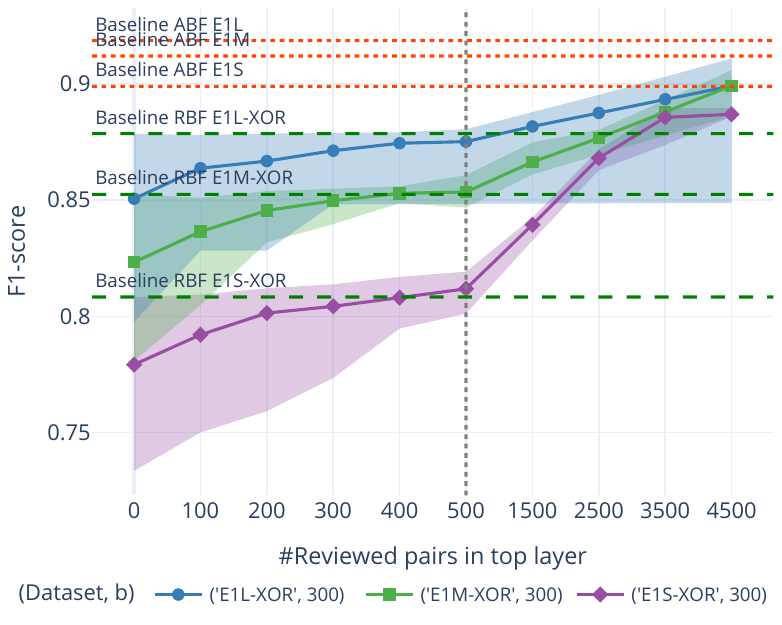}
\caption{Linkage quality development for datasets E1S, E1M and E1L with XOR hardened CLK encodings, fixed $b = 300$ and $err = 0.2$.}
 \label{fig:quality-223x-bX-xor-300}
\end{figure}
\end{minipage}
\vspace{-0.8cm}
\end{wrapfigure}
\subsubsectionsubst{BF Hardening}
The experimental outcomes based on $E_R$ with XOR hardening (see Fig.~\ref{fig:quality-223x-bX-xor-300}) are comparable in terms of the average F1 scores, which improve by $0.108$ (E1S-XOR), $0.076$ (E1M-XOR) and $0.049$ (E1L-XOR). However, the linkage quality is in general slightly lower as a trade-off for the improved privacy of the encodings, which is illustrated, e.g., for E1M-XOR compared to E1M by decreases of \textit{G} from $0.224$ to $0.095$ and \textit{JSD} from $0.172$ to $0.079$.

\subsubsectionsubst{Datasets E2S and E2M}
Finally, we study the results for datasets E2S and E2M (see Fig.~\ref{fig:quality-22xx-bX} (right)).
In general, the linkage quality is lower due to the higher dissimilarity of duplicates.
The average F1 score is raised by $0.109$ (E2S, $b=300$) and $0.048$ (E2M, $b=300$), which is comparable to the E1 datasets.
However, the progress is less stable, leading to a larger range of outcomes and for some runs even quality losses (E2S, $b=200$).

For most of the datasets the respective reference results solely using (non-keyed) ABF encodings (orange lines) are not reached on average.
Please note, though, that the baseline results are reported for an optimized classification threshold.
Nonetheless, the distances of the final average F1 scores are rather small: $-0.008$ (E1S), $-0.003$ (E1M), $-0.008$ (E1L), $0.012$ (E2S) and $-0.026$ (E2M) (all results for $b=300$ and $err=0.2$).

\subsubsectionsubst{Privacy}
Table~\ref{tab:abf-privacy-measures} compares the privacy measures of these ABF baselines with our approach using \textit{keyed} attribute-level Bloom filters. The number of available encodings in $L_A$ is decreased from up to $100.000$ (total number of records in the dataset) to up to $9.000$ (two times the number of record pair review requests by $L_R$). For some attributes, in particular middle name and place of birth, the counts are lower due to missing values.
Both Bloom filter privacy measures are heavily reduced from $0.147 - 0.388$ (\textit{G}) and $0.114 - 0.299$ (\textit{JSD}) to at most $0.01$. For attributes with a high variety of values, in particular name components, the decreasements are the lowest. The largest improvements are achieved for the year of birth attribute, likely due to very frequent bigrams such as '19' which are reflected in having very frequent corresponding bit positions in conventional ABF. This illustrates the privacy-enhancing effect of the pair-specific hashing secrets in our protocol.

\begin{table}
  \caption{Privacy measures in $L_A$ for Keyed Attribute-level Bloom filter (KABF) compared to the baseline approach using only ABF-based linkage for linkage of E1M. The given values for KABF are the mean over all experimental runs for that dataset.
  }
  \label{tab:abf-privacy-measures}
  \setlength{\tabcolsep}{3.9pt}
  \begin{tabular}{llrrrrrrr}
    \toprule
     & Enc. & FN & MN & LN & YOB & CITY & ZIP & POB\\
    \midrule
    \multirow{2}{*}{Number of encoded attributes} & ABF & $100k$ & $92.2k$ & $100k$& $100k$ & $100k$ & $99.9k$ & $80.8k$\\
    & KABF & $9k$ & $8.4k$ & $9k$& $9k$ & $9k$ & $9.0k$ & $7.6k$\\
    \multirow{2}{*}{Gini coefficient (\textit{G})} & ABF & $0.171$ & $0.157$ & $0.147$& $0.388$ & $0.212$ & $0.330$ & $0.373$\\
    & KABF & $0.010$ & $0.009$ & $0.010$& $0.010$ & $0.009$ & $0.009$ & $0.010$\\
    \multirow{2}{*}{Jensen-Shannon diverg. (\textit{JSD})} & ABF & $0.132$ & $0.121$ & $0.114$& $0.299$ & $0.162$ & $0.251$ & $0.296$\\
    & KABF & $0.008$ & $0.007$ & $0.007$& $0.008$ & $0.007$ & $0.007$ & $0.008$\\
  \bottomrule
\end{tabular}
\end{table}

Fig.~\ref{fig:privacy-measures} shows the distributions of the privacy measures in $L_C$ for datasets E1M and E2M based on the final states of the runs with $b=200$ and $err=0.2$.
Each measure is computed for three attribute selection methods: \textit{No restrictions} refers to the baseline where all attributes of uncertain pairs in $L_A$ are provided to $L_C$.
In \textit{No equal attributes} the data owners are asked to provide all attributes whose similarity is below $1$, whereas in the last setting, an additional filter is applied to exclude very dissimilar attributes ($sim<0.4$) as well.
In the first setting, the results represent the availability of the attributes in the plain dataset. With stronger restrictions, the share of attributes that have been requested and provided decreases for most attributes. While all attributes of E1M apart from MN and POB are available for (nearly) all records in the unrestricted setting, the average sharing rate is reduced to a high (FN, YOB) or moderate (LN, CITY) degree using the proposed selection strategies. The ZIP code, however, is still available in plaintext for most records (median $\approx 83\%$).
For E2M the availability of the location-related attributes is even higher, close to $100\%$ for the ZIP code.
While the median scores are otherwise comparable to E1M, the results show a lower variance.
The KAPR score for both datasets is reduced from above $0.9$ to below $0.4$ and $0.2$ using the two filtering methods.

\def\pbpplotwith{0.48\linewidth}
\begin{figure}[ht]
\begin{minipage}[b]{\pbpplotwith}
\centering
\includegraphics[width=\textwidth]{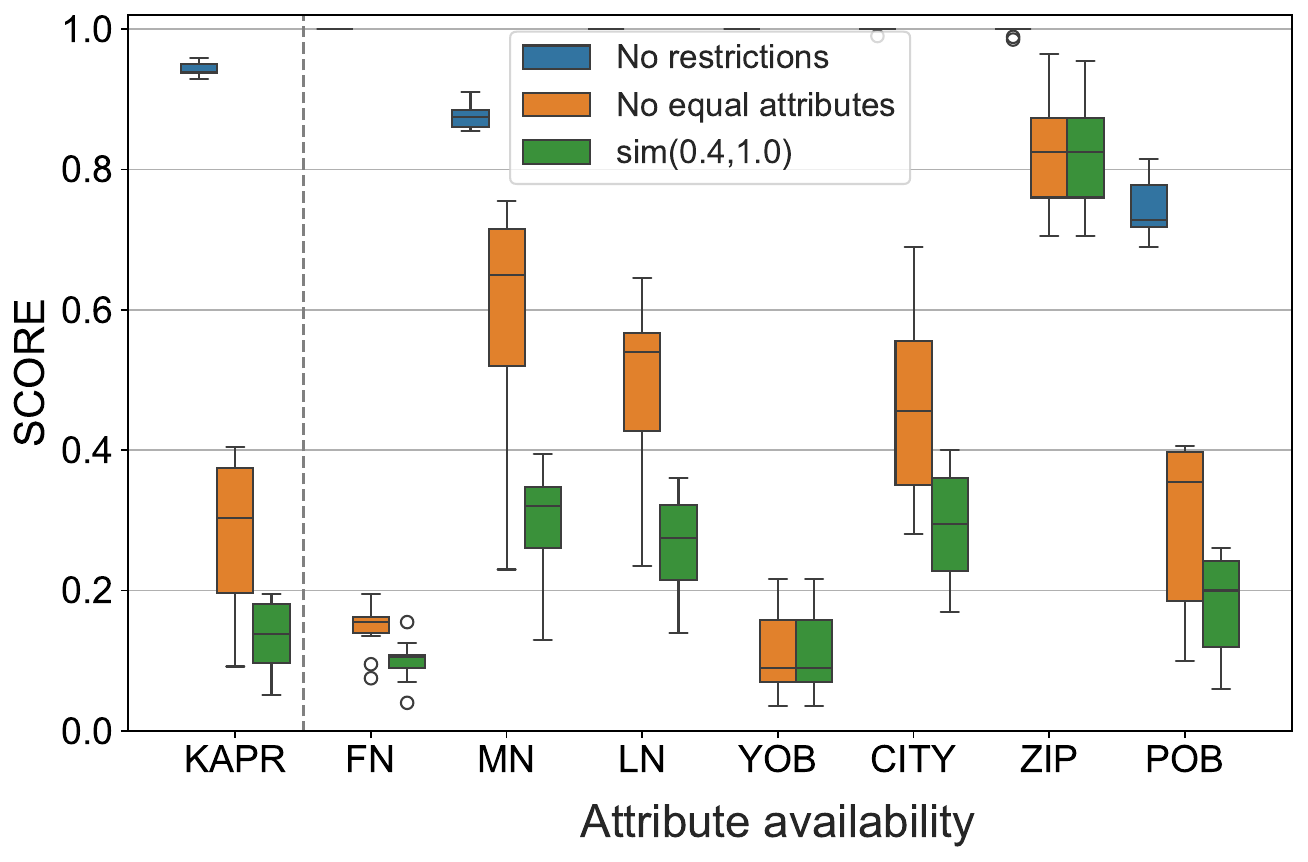}
\end{minipage}
\hspace{0.5cm}
\begin{minipage}[b]{\pbpplotwith}
\centering
\includegraphics[width=\textwidth]{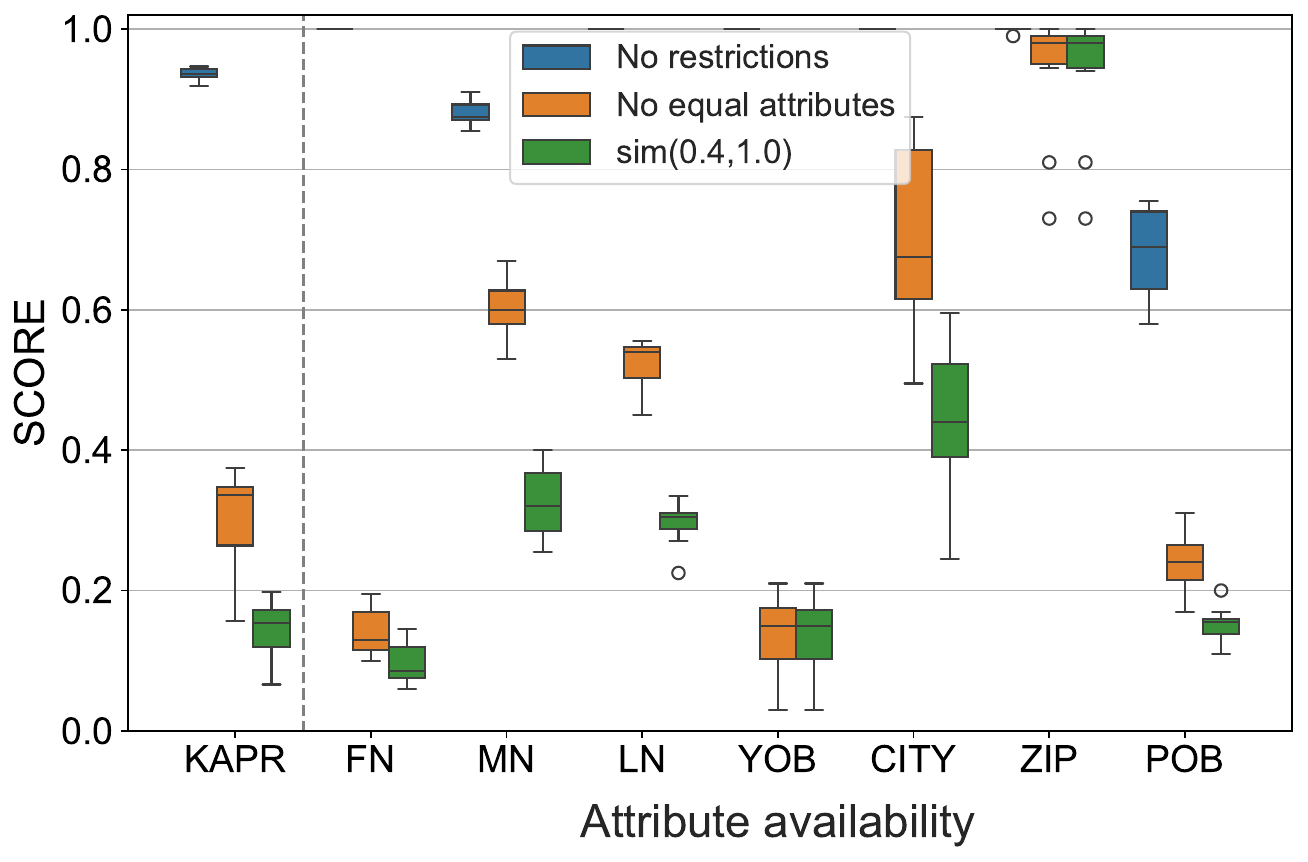}
\end{minipage}
 \caption{Comparison of the privacy measures in $L_C$ for datasets E1M (left) and E2M (right) using different attribute selection strategies. All measures to the right of KAPR refer to the availability of the respective attribute in this layer.}
 \label{fig:privacy-measures}
\end{figure}

%% file: 5_discussion.tex
\section{Discussion}
In general, the results show that the proposed protocol achieves its aim of improving the linkage quality with a limited amount of manual labeling.

\subsubsectionsubst{Quality}
While higher error rates of the masked clerical review naturally are detrimental to the overall performance, it can be observed from Fig.~\ref{fig:quality-2232-eX-bX} (right) and Fig.~\ref{fig:quality-22xx-bX} (left) 
that protocol runs with increased budgets achieve F1 score improvements by approximately $5-10\%$ on average.
Both aspects, clerical reviews as well as the model updates, contribute to that. The tuned threshold of \modelRL is largely responsible for improvements in early iterations with few reviewed instances.
F1 scores above the green reference lines represent results that are beyond the optimal initial threshold configuration and therefore could not have been achieved without clerical review based on additional data in $L_A$ and $L_C$.
In addition, the range of possible outcomes is reduced substantially. This means that data custodians can expect a high linkage quality, less depending on the initial classification threshold choice.

This also applies to the E2 dataset variants in principle, however, a higher budget is required for stable results. As the oracle error rate is identical to the E1 experiments, the cause must be the training of \modelAL and \modelRL. The high availability of the residence attributes and the low variance of other attributes in $L_C$ (Fig.~\ref{fig:privacy-measures}) indicates that the reviewed pairs are fairly similar, leading to a biased and poor \modelAL. In consequence, the performance of \modelRL also deteriorates.

In our setup, the warm-up phase with $\mathit{AL}_A$ in the first five iterations determines the performance of \modelAL, because the training will be stopped as soon as the clerical review budget $b$ is reached.
\modelAL serves as the bridge between the more secure comparison of the majority of records based on record-level encodings and the masked clerical review for a small set of uncertain match candidates.
For this functionality, it is not necessary that \modelAL performs better than \modelRL. In fact, the protocol would even work without the model when $(g,p)$ of \modelRL is reused in this layer. In that scenario, the computed attribute similarities based on the attribute-level encodings would merely serve to restrict the requested plaintext data for the subsequent clerical review to non-equal attributes.

In order to achieve higher quality improvements, an increased number of reviews is required, because \modelRL cannot be improved much due to its privacy-induced simplicity with a single feature. However, the number of manual reviews should be kept as low as possible. Therefore, it is important that \modelAL has a high predictive performance to provide the majority of corrected links.
In our evaluation setup we used a fixed budget of $4500$ reviews by $L_A$, analogous to the clerical budget $b$ in $L_C$. In principle, additional reviews are possible and may raise the linkage quality even further. However, the likelihood that pairs having similarities with larger distance to the threshold in \modelRL are wrongly classified declines, deteriorating the ratio of corrected to reviewed links.

A possible approach to improve our setup could be a more sophisticated query strategy in $LS_A$ based on the available feature vector instead of the probability score only.
Furthermore, it could be beneficial using a pre-trained \modelAL instead of bootstrapping it based on the initial predictions of \modelRL.
Such an approach requires a sufficiently similar reference dataset with a given ground truth for training though.

The ABF baseline results are higher than the RBF baselines as they make use of attribute weights and weight redistribution in case of missing values, even with weights determined on an independent dataset (German cancer registry).
The experiments show that our protocol achieves only slightly lower linkage quality compared to this baseline linkage approach. However, the risk of reidentification attacks based on frequent patterns in the underlying attribute-level encodings is greatly reduced, as illustrated by Table~\ref{tab:abf-privacy-measures}.

\subsubsectionsubst{Privacy}
The majority of reviews is handled automatically by the intermediate layer without plaintext access while only a small fraction is reviewed manually based on partial display. After the warm-up, $\frac{b}{500} = 20/40/60\%$ of the uncertain pairs from the initial layer have been reviewed using PPCR, the remainder by \modelAL. At the end, the proportion is reduced to $\frac{b}{4500} = 2/4/7\%$.
We studied basic filtering approaches for lowering the number of requested plaintext attributes based on the attribute similarity. The observed reductions vary between the attributes. In particular, the effect for the numerical ZIP code attribute is comparatively low, because it has a high similarity on average due to the records being from the same US state. As a consequence, it is requested for a high share of pairs. This problem may be addressed in future work by using a dynamic data-driven approach where the lower bound is determined by the average similarity of each attribute in $L_A$.

The KAPR scores are drastically decreased due to the attribute selection. Fewer persons can be uniquely reidentified based on the remaining attributes. This is also illustrated by the observation that the availability is particularly reduced for strongly identifying attributes like name and year of birth.
Nevertheless, malicious reviewers could still target certain persons, e.g., with rare attribute combinations. The data owners may implement their own selection rules to protect their high profile records.

We want to point out that although our protocol was described and evaluated using Bloom filter encodings due to their popularity, any perturbation technique is applicable that supports record-specific keys and approximate similarity computation. As the experimental results using hardened Bloom filter encodings (see Fig.~\ref{fig:quality-223x-bX-xor-300}) suggest, the overall linkage quality might be lower though, despite comparable improvements.

%% file: 6_conclusion.tex
\section{Conclusion}
By their very nature, privacy-preserving classification problems are difficult to parameterize in practice due to the lack of labeled training data, e.g., from clerical review.
In privacy-preserving record linkage, record-level Bloom filter encodings are frequently used to improve resistance against reidentification attacks but suffer from potential quality issues and the need for a well-selected classification threshold.
We presented an active learning based protocol that enables increased and more stable linkage quality using multiple layers of clerical review. Merely a low number of masked manual reviews is required because the majority of uncertain pairs can be classified automatically with higher accuracy based on attribute-level features with the trained models.
Sharing additional information with the linkage unit increases the risk of reidentification. We therefore employ record-specific salting and attribute selection to hamper such attacks.
Furthermore, data owners still remain in full control of the information they are willing to share for each record.

We sincerely hope that the increased reliability of the linkage outcome leads to a wider adoption of PPRL methods in practical applications.
Our proposed multi-layer clerical review is also applicable in incremental linkage, e.g., for pseudonymous patient registries, and would allow monitoring of the linkage quality over time.
In future work, we will investigate whether the suitability of record-level encoding parameters can be assessed based on the ability of the record-level model \modelRL to reproduce labels from the attribute-level model \modelAL, and how optimal weights can be determined based on attribute agreements of selected match candidates.